\documentclass[article]{aa}
\usepackage{graphicx}
\begin{document}

\def\SBS{SBS~1520+530}
\def\hyp{{\it Hyperz} }   
\def\ho{H$_0$}
\def\hirac{HiRAC }
\def\sex{{\it SExtractor} }
\def\kms{km.s$^{-1}$ }
\def\kmsMpc{km.s$^{-1}$.Mpc$^{-1}$ }

   \title{The lensing system towards the doubly imaged quasar
SBS~1520+530\thanks{Based on observations obtained at the 2.56m Nordic
Optical Telescope (La Palma, Spain) and with the Hubble Space
Telescope, operated by NASA.}}
    
\subtitle{}

\author{C. Faure\inst{1,2}\and F. Courbin\inst{3,4} \and J.P. Kneib
\inst{2} \and D. Alloin\inst{1} \and M. Bolzonella \inst{2,5} \and
I. Burud \inst{6,4}}
   
   \offprints{C\'ecile Faure: cfaure@eso.org}

   \institute{
         European Southern Observatory, 
Alonso de Cordova 3107, Vitacura, Casilla 19001, Santiago 19, Chile
        \and 
Observatoire Midi$-$Pyr\'en\'ees, UMR 5572, 14 avenue
        Edouard Belin, 31400 Toulouse, France
       \and
       Universitad Cat\'olica de Chile, 
Departamento de Astronomia y Astrofisica, Casilla 306, Santiago 22, Chile
        \and 
Institut d'Astrophysique et de G\'eophyique de Li\`ege, 
Avenue de Cointe 5, B-4000 Li\`ege, Belgium
        \and
Istituto di Fisica Cosmica "G.Occhialini", Milano, Italy 
        \and 
Space Telescope Science Institute,  
3700 San Martin Drive,   Baltimore, MD 21218 , USA
}

\date{Submitted: November 8, 2001}

   \abstract{The gravitational potential responsible for the lensing
effect in \SBS\, is studied over length scales from a few arc-seconds
to a few arc-minutes. For this purpose, we use sharply deconvolved Hubble Space Telescope images in the optical
and near-IR, in combination with ground based optical data obtained over a wider field-of-view. In particular, we have carried out a 
multi-color analysis in order to identify groups or
clusters of galaxies along the line of sight. Photometric redshifts
are measured for 139 galaxies unveiling significant excesses of
galaxies  1.0\arcmin\, NW and 1.7\arcmin\, SW of the main lensing
galaxy. The photometric redshift inferred both for the main lensing
galaxy and for the galaxy concentrations is
 z=$0.9^{+0.10}_{-0.25}$. 
This is in rough agreement with the measured spectroscopic redshift of the main lensing galaxy, z=$0.71$ (Burud et al. 2002), suggesting
 that it is part of a larger group or
cluster. We investigate the impact of including the galaxy cluster, first on the modelling
of the lensing system, and second on the expected time--delay between the 
two quasar
images.   }

\titlerunning{The lensing system toward \SBS}
   
 \maketitle
%
\section{Introduction}
 
The study of multiply imaged quasars is one of the most promising way
to measure cosmological parameters such as the Hubble parameter \ho\, (Refsdal 1964, Blandford \& Narayan 1992). Indeed, \ho\, is 
related to two observables: the time--delay between the light curves of
the quasar images and the mass distribution in the lens.  Until
recently, simple mass distributions were used to model multiply imaged
quasars. However, the improvement of observing techniques has led to
the discovery that most image configurations require more complex
models, involving, for example, an external shear (Keeton, Kochanek
\& Seljak 1997).  This small but significant external perturbation to
the main lensing potential often corresponds to the presence of groups
or even clusters of galaxies along the line of sight (see for example
Keeton \& Kochanek 1997, Burud et al. 1998, Morgan et al. 2001, Fassnacht \& Lubin 2001). These additional structures must be taken into account
in order to properly model the mass distribution and accurately
convert the observed time--delay into \ho.\\ 
\SBS\, is a doubly imaged quasar at z=$1.855$, with an angular
separation of 1.568\arcsec\,
($\alpha_{2000}$=15$h$\,21$m$\,44.83$s$\,,
$\delta_{2000}$=+52$\degr$\,54\arcmin\,48.6\arcsec). It was discovered
by Chavushyan et al. (1997) in the course of the Second Byurakan
Survey (Markarian \& Stepanian, 1983).
 Crampton et al.  (1998) identified the lensing galaxy from near-IR
adaptive optics images and proposed a simple lens model including
exclusively the main lensing galaxy.  They concluded however that an
additional external shear was mandatory to model properly the
system.\\ Deep ground--based and HST data are used in the present
paper to map the mass distribution along the line of sight to SBS
1520+530. We then investigate the source of external shear and propose
a multi--components model for the total lensing potential, where the
main lensing galaxy is a member of a larger group or cluster of
galaxies.


\begin{figure*}[hbtp]
\begin{center}
\includegraphics[width=18cm]{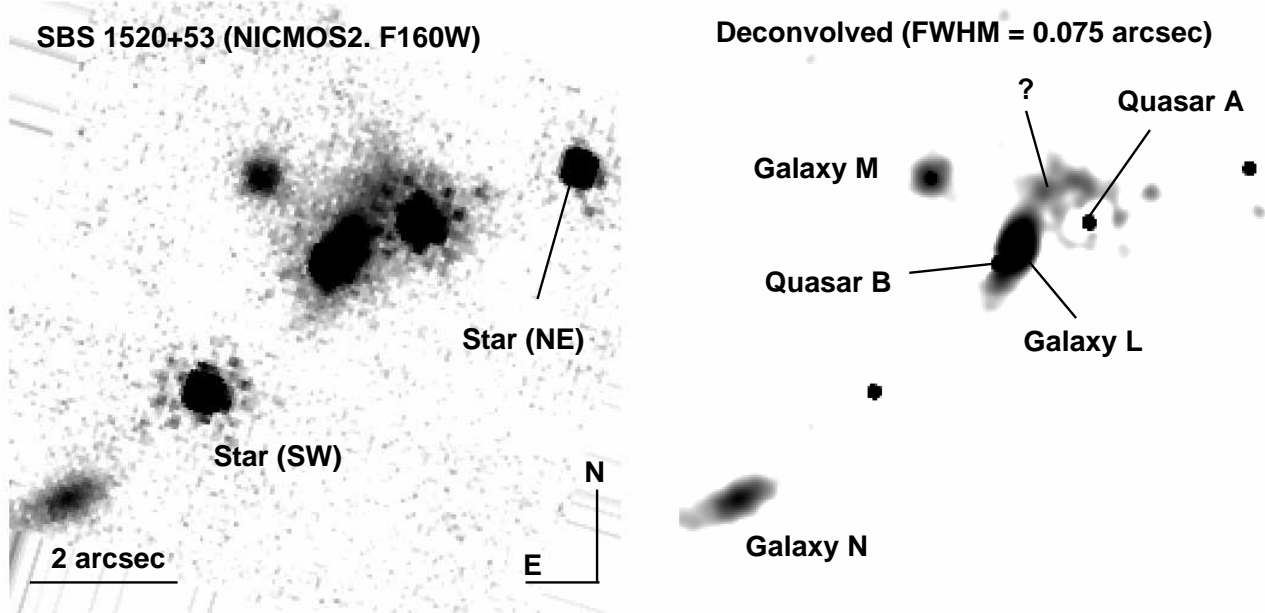}
\caption{\label{fig1} Simultaneous deconvolution of the four
HST/NICMOS2 frames obtained of \SBS. The left panel shows a standard
combination of the data. The right panel is the result of the
deconvolution, with a final resolution of 0.075\arcsec. The pixel size
on this 19\arcsec wide image is 0.0375\arcsec. North is to the top,
East is to the left. The labels follow the description in section \ref{result}.}
\end{center}
\end{figure*}

\section{Observations - data reduction}

\subsection{Ground--based observations}

The ground--based optical data consist of two sets of images obtained
during two observing runs, with the 2.56m Nordic Optical Telescope
(NOT) on the island of La Palma, Spain.

\SBS\, was observed with the High Resolution Adaptive Camera (HiRAC)
on the night of February 25, 1998 under photometric conditions and
average seeing between 0.8\arcsec\, and 1.2\arcsec. The total exposure
times were 90 min in $I$, 40 min in $R$ and 30 min in $V$. With a
pixel size of 0.11\arcsec, the useful field--of--view of the 2102
$\times$ 2052 detector is about 3.5\arcmin\, on a side. The images
were reduced using sky flat-fields. Strong fringes (10\%) are
affecting the $I$-band images, but many dithered short exposures were
combined into a high signal-to-noise map of the fringes which was then
subtracted from the data. This procedure leads to  final images with sky
subtraction accurate to the percent level.

Additional images were obtained on the night of February 16, 1998,
with the Andalusia Faint Object Spectrograph and Camera (AlFOSC). The
pixel size is 0.188\arcsec\, on this instrument, which sets the field--of--view to 6.5\arcmin. The observations were taken under 0.9\arcsec\,
seeing and through thin cirrus. Total exposure times were 25 min in
$R$ and 30 min in $V$. These images were combined with the \hirac
(flux calibrated) data to obtain a deeper dataset but were also used
to look for overdensities of galaxies over a wider field--of--view
than that of the \hirac data.

\subsection{Hubble Space Telescope observations}

Archival data of \SBS\, were used to derive the precise astrometry of
the system in the immediate vicinity of the lens. These data are part
of a much larger public survey of gravitational lenses (PI: E. Falco)
known as the CfA-Arizona Space Telescope LEns Survey
(CASTLES). NICMOS2 $H$-band (F160W) data were taken on July 20, 1998,
with a total exposure time of 47 min. At the date of observations, the
NICMOS2 pixel size was 0.0759\arcsec $\times$ 0.0753\arcsec.

Optical data were obtained with WFPC2 through the F814W filter (August 22, 1999), with a total exposure time of 27 min and through the F555W
filter (November 11, 1999), for a total exposure time of 35 min.

\section{Image deconvolution}

\begin{table*}[t]
\begin{center}
\renewcommand{\arraystretch}{1.4}
\begin{tabular}{|l|c|c|c|c|c|c|c|}
\hline
               & F555W & F814W &I& F160W & $\Delta ~ \alpha$ ('')&
$\Delta  \delta$ ('') & z\\
\hline
\hline
Date   &11/11/1999 &22/08/1999 &25/04/1998 & 22/07/98 & /&/&/ \\
\hline
A  &18.86$\pm0.04$ &17.91$\pm0.03$ &17.68$\pm0.02$ &17.09$\pm0.02$  & 0&0 &1.855 \\
B  &19.31$\pm0.06$ &18.83$\pm0.03$ & 18.49$\pm0.02$&18.17$\pm0.03$&-1.426$\pm0.006$ &-0.652$\pm0.002$ &1.855\\
\hline
$\frac{Flux(B)}{Flux(A)}$ & 0.576 &0.436  &0.476 &0.465 & / &/ &/ \\
\hline
gal L &  24.8$\pm0.5$  & 21.8$\pm0.2$ & 21.17$\pm0.10$ & 18.75 $\pm0.10$  &-1.130$\pm$0.005 &-0.387$\pm$0.001 &0.88 $^{+0.02}_{-0.26}$ \\
\hline
gal M &    24.5$\pm0.3$  & 22.6$\pm$0.2 & 21.84$\pm$0.10&20.03 $\pm0.10$  & -2.495$\pm$0.005 & 0.690 $\pm$0.005&  0.89$^{+0.06}_{-0.06}$   \\
\hline
\end{tabular}
\caption{\label{tab1} Information about the doubly imaged quasar and the lensing galaxies. The lines successively provide: (1) date of the observations, (2) magnitudes of the A and B images of the quasar (see designation in Fig. \ref{fig1}), coordinates relative to A, and spectroscopic redshift (Chavushyan et al. 1997), (3) flux ratio of the two quasar images, (4) magnitudes, coordinates relative to A and photometric redshift for galaxy L (with error bars at 1$\sigma$), (5) similar data for galaxy M.}
\end{center}
\end{table*}

\subsection{Image processing}

Several dithered images are available for each filter: 3 in F555W, 4
in F814W and 4 in F160W. These images were combined after cosmic rays
and bad pixels rejection using standard IRAF procedures and were
subsequently deconvolved using the ``MCS'' deconvolution algorithm
(Magain et al. 1998). The algorithm can either deconvolve one single
frame or deconvolve simultaneously several frames on the same object,
combining them at the same time into one deep sharp frame. The
algorithm also decomposes the data into a sum of point sources (here
the quasar images and nearby stars) and extended sources (the lensing
galaxy). One can therefore derive the photometry of the lens,
decontaminated from the quasar light.

We deconvolved the data in two ways. The F555W and F814W data were
deconvolved only after combining the individual images, because of poor
signal-to-noise ratio. For the more numerous and deeper F160W images, the
individual data frames were {\it simultaneously deconvolved} in order
to produce a high quality image of the system. This process is the
same as described in Courbin et al. (1998) with a slight deviation from
the standard application of this technique. In the standard
application, each frame is seen as the convolution of a unique ideal
frame with the PSF associated to each image in the data set. In the
present case, the PSF available in the data (stars ``SE'' and ``NW'' in Fig.~1) are faint compared to
the quasar images and we chose to compute
one single PSF for all frames. This PSF is constructed from all stars
available in the data set, i.e. two stars per frame, times four
frames. This has the advantage of providing a high signal-to-noise PSF
with very good rejection of cosmic rays and bad pixels but it also
implicitly assumes the temporal stability of the PSF over the period
of observation.  The quality of the residuals after deconvolution (see Courbin et al. 1998 for more details) remains good, indicative
that our assumption is actually correct.


\subsection{Results}\label{result}

The result of the simultaneous deconvolution of the F160W images is
displayed in Fig.~\ref{fig1}. Objects are labelled as in Crampton et
al. (1998). The two PSF stars used for
the deconvolution are the ones labelled ``SE'' and ``NW''.  Note the
elongated lensing galaxy L appearing between the two quasar images. Two
other objects are also visible in the immediate vicinity of the lensed
source (labelled galaxies M and N in Fig.~\ref{fig1}). 
Although at the confidence limit after the deconvolution process, the extended plumes to the NW of the lensing galaxy L might trace a prior interaction with galaxy M. However we shall ignore this feature in the subsequent modelling.
Table~\ref{tab1} summarizes the main properties of the quasar images
and of the lensing galaxy.  As the lensing galaxy appears to be
essentially elliptical, models are particularly sensitive to its
orientation and ellipticity: hence particular attention was paid to
the precise determination of these parameters. We used the {\it
Im2shape} software (Bridle et al. 2002) to determine accurate values of
the PA and of the ellipticity of the galaxies L and M. We measure an
ellipticity of 0.50$\pm$0.02 for galaxy L, the ellipticity being
defined as $\epsilon=(a^2-b^2)/(a^2+b^2)$. Its major axis is oriented
at PA$=-23^o\pm 7^o$. Galaxy M is located at 2.56\arcsec\, to the NE
of the quasar image A and is likely to play some role in the lens
modelling. It is almost circular ($\epsilon$=0.04$\pm$0.02, PA$=-40^o\pm
10^o$). Galaxy N is too far away from the main lensing galaxy L
($\sim$7\arcsec) to play an important role in the modelling,
at least if realistic masses are considered. The coordinates
of the quasar image B, of the lensing galaxy L and of galaxy M, relative to the quasar image A, are
provided in Table~\ref{tab1}. Photometric redshifts have been derived
for the different objects in the immediate vicinity of \SBS.
 These redshifts were estimated with {\it Hyperz}\footnote{The original code and user's
manual are available at http://webast.ast.obs-mip.fr/hyperz} (Bolzonella et al. 2000), as were those of the intervening galaxy
cluster/group (see Section \ref{photz} for complete details). Results from {\it Hyperz} consist in two
different redshift estimates. The first redshift corresponds to the best fit (smallest $\chi ^2$) obtained after testing a wide  variety
of galaxy templates at different redshifts. It is called the 
``photometric redshift'', z$_{phot}$. The second redshift corresponds to a  
weighted mean redshift z$_{wm}$, representative of the ``most
frequent'' redshift allocated to the galaxy, while 
testing a family of galaxy templates over a range of redshifts. The latter redshift, z$_{wm}$, is less
affected by degeneracies due to redundant spectral features in galaxy
templates. We obtain, for galaxy M, z$_{phot}$=0 .89 $^{+0.02}_{-0.26}$ and z$_{wm}$=0.57, and for the main lensing galaxy L, z$_{phot}$0 .88 $^{+0.02}_{-0.26}$ and z$_{wm}$=0.71. At the time we were performing this analysis, a
spectroscopic redshift of galaxy L was measured, z=0.71 (Burud
et al. 2002). These three values are consistent within the error bars.

\begin{figure*}[t]
\centering
\caption{\label{alfosc}The AlFOSC $R$-band image (5.7\arcmin $\times$
6.0\arcmin) around \SBS. The isodensity contours outline regions with
constant number of galaxies per square arc-minute. Notice the concentrations of galaxies to the NW and SW of the lensed quasar (indicated by an arrow).}
\end{figure*}
\begin{figure}[h]
\begin{center}
\caption{\label{fig5bis} Similarly to Fig. \ref{alfosc}, isodensity contours determined for the
2.8\arcmin\, $\times$ 2.7\arcmin\, \hirac field (I-band) around SBS
1520+530. The edges of the overdensity of galaxies already seen in the
AlFOSC image, are also detected in the NW and SW corners of the
frame.}
\end{center}
\end{figure}
\begin{figure}
\begin{center}
\includegraphics[width=5.8cm]{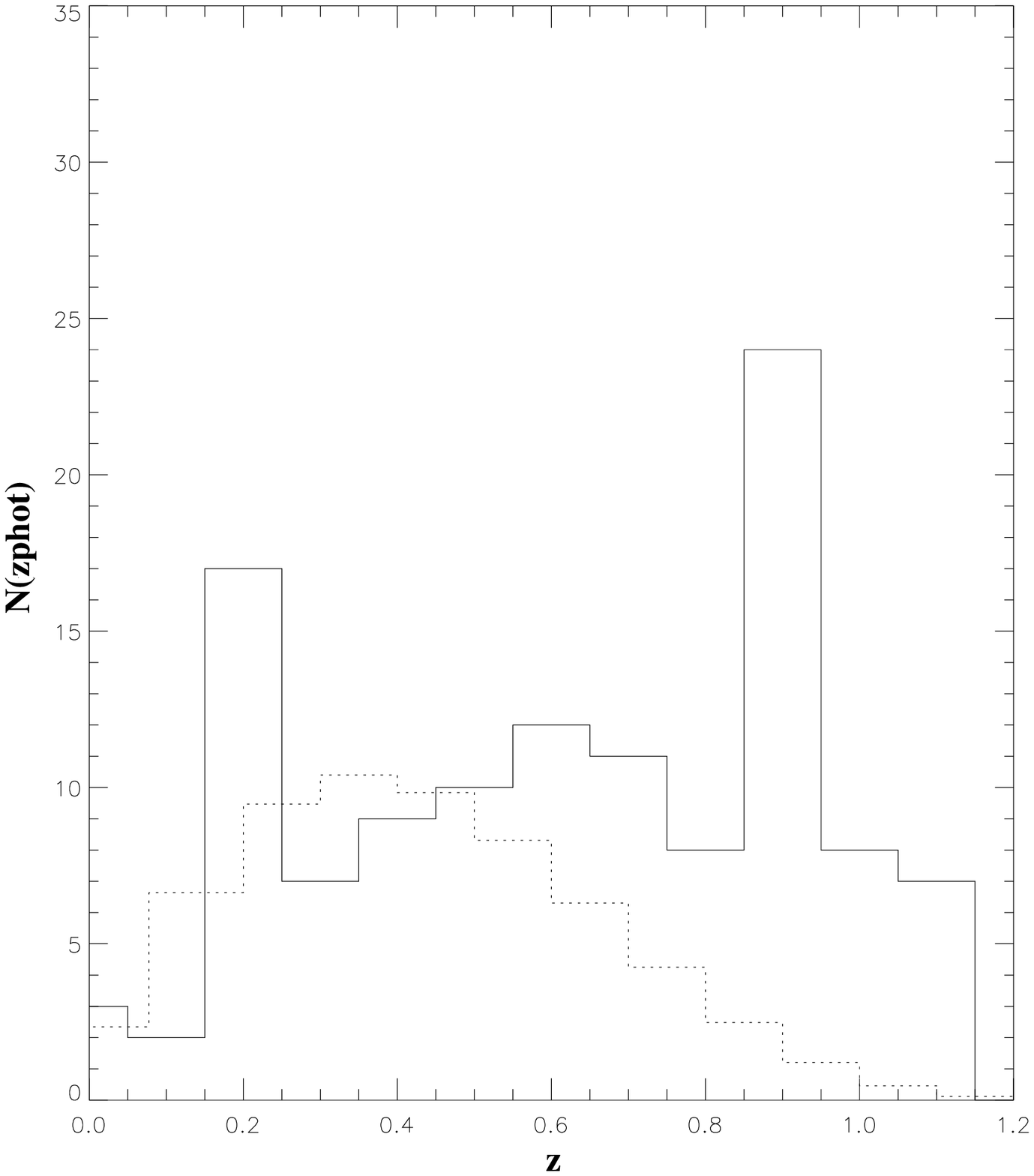}
\includegraphics[width=5.8cm]{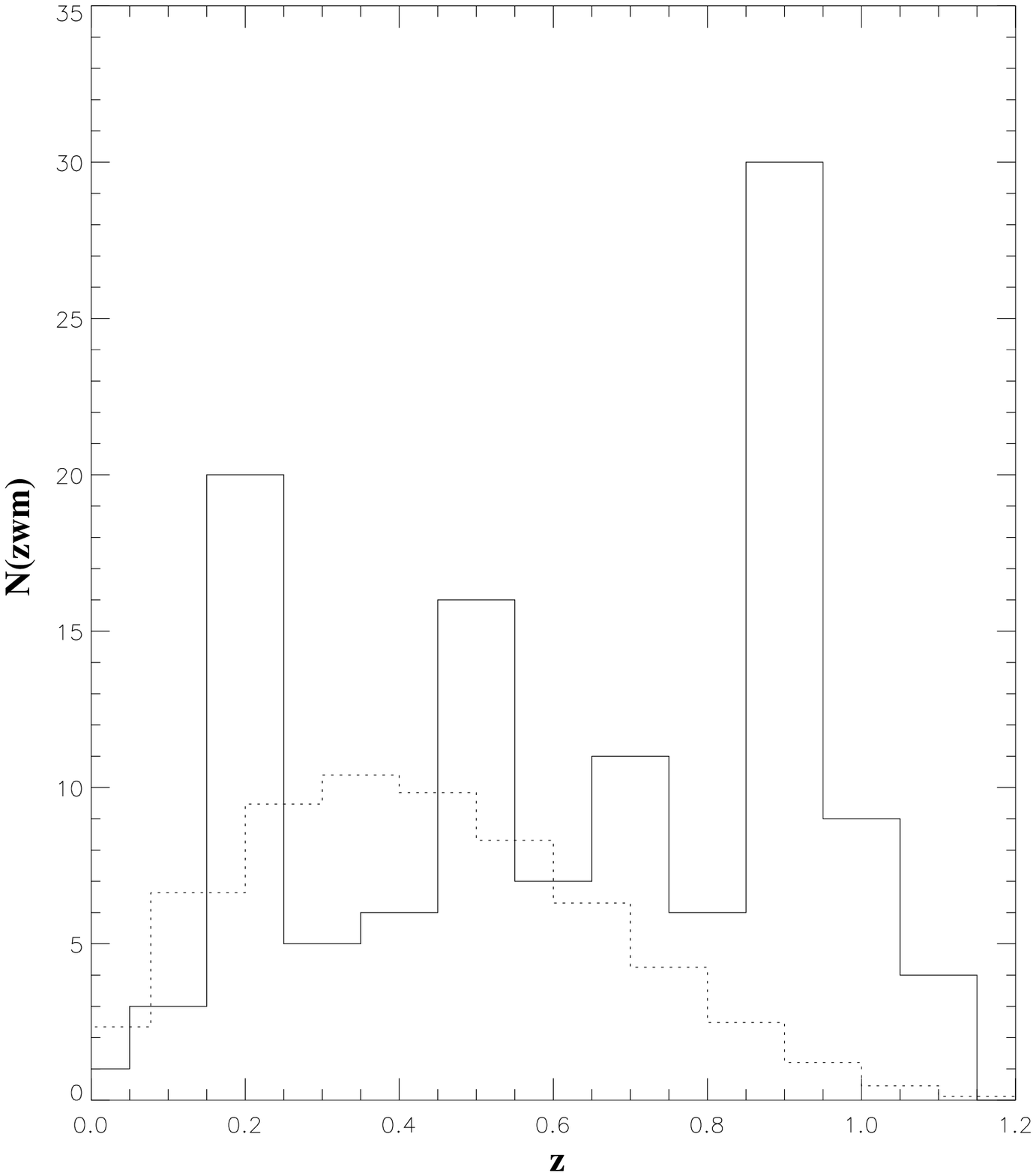}
\caption{\label{fig5} Relative distribution of galaxies with two
different redshift estimators, for the \hirac field--of--view around
SBS1520+530 (solid line). {\bf Top panel}: distribution for the
photometric redshifts (z$_{phot}$). {\bf Bottom panel}: distribution
for the weighted mean redshifts (z$_{wm}$). In each panel we have also
plotted the distribution expected for a reference field of the same
size and containing no cluster/group (dashed line).}
\end{center}
\end{figure}

\section{Detection of an intervening galaxy cluster/group}\label{sec3}
\subsection{Photometry of the field: an overdensity 
of galaxies}\label{section3}

The photometry of all galaxies detected in the field--of--view near \SBS\,
was performed using the {\it SExtractor} 1.2 software (Bertin \& Arnouts 1996). No obvious defects were present on the detector chip. Yet the
central part of the field (20\arcsec\, in diameter) was masked, as it
is saturated by a very bright star.

The presence of an intervening galaxy cluster/group along the line of
sight to SBS~1520+530 was first investigated on the wide field
dataset provided by AlFOSC. In the $R$-band, we have detected 620
galaxies over a 6.5\arcmin\, field--of--view after removal of the stellar objects. A stellar object was defined as
an object brighter than $R=19$, with a FWHM smaller than 1.0\arcsec\,
(the seeing being 0.9\arcsec\,) and with elongation (defined as $a/b$)
smaller than 1.1.  Isodensity contours were obtained from the AlFOSC
image, for all galaxies in the range $22<R<25$. This eliminates low
redshift ``normal'' galaxies but may still include some contamination by
faint dwarf galaxies which cannot be identified using only $V$, $R$ and $I$
photometry. Following these criteria we have displayed the $R$-band
image in Fig.~\ref{alfosc} where a significant overdensity of
galaxies is seen close to the centre of the field. The projected
density of galaxies is a factor two larger in the densest parts than in the rest of the field. The shape of the overdensity is
irregular and even double-peaked, with one peak located at 1.0\arcmin\, NW of
\SBS\, and a second peak at 1.7\arcmin\, to the SW.  These
overdensities also appear in the smaller field--of--view of the \hirac
$I$-band image, applying similar selection criteria, i.e. rejecting
the stellar objects and considering only the objects with $21<I<24$ (see
Fig.~\ref{fig5bis}).

\begin{figure*}
\centering
\caption{\label{couleur} Composite image corresponding to the \hirac
field--of--view (2.8\arcmin\, on a side taking into account dithering).
The $V$, $R$ and $I$ images have been used to build this
``true-color'' image. In this image, violet galaxies (large blue circles)
are allocated redshifts in the range 0.1$<$z$<$0.3 by {\it Hyperz} while
the red ones (small yellow circles) are in the higher redshift bin
0.7$<$z$<$1.0, as the main lensing galaxy.}
\end{figure*}

\subsection{Photometric redshift analysis}\label{photz}

Multi--color information is available in a 2.8\arcmin\, wide field
around \SBS\, which corresponds to the \hirac field--of--view. This is
sufficient to recognise the concentration of galaxies detected in the AlFOSC
image and to infer photometric redshifts, based on their $V$, $R$ and
$I$ magnitudes. A total number of 139 galaxies are detected in the three
filters simultaneously. The magnitudes and their error bars are used
as the input to {\it Hyperz} (Bolzonella et al. 2000), in order to estimate redshifts.  {\it
Hyperz} makes a comparison between the photometric spectral energy
distribution (SED) of the observed galaxy and those obtained from a
set of reference template spectra, using the same photometric system. Reddening
is also taken into account: we used the Calzetti law (Calzetti et
al. 2000). The limiting magnitudes, corresponding to $\sim$80\%
completeness, are 25 in $V$, 24 in $R$, and 23 in $I$. This allows to
detect objects up to z$\sim$1.2. The entire library of
template spectra proposed in {\it Hyperz} has been considered, encompassing a variety of star formation histories. 

The most likely redshift is estimated for each galaxy through a
$\chi^2$ minimization. As the distribution of $\chi^2$  versus redshift is
not symmetrical around the best redshift value, the weighted mean
redshift z$_{wm}$  can also be a relevant and interesting parameter.
In Fig. \ref{fig5}, we have plotted the relative distribution of
galaxies as a function of redshift, using both the photometric redshift
z$_{phot}$ (upper plot) and the weighted mean redshift z$_{wm}$ (lower
plot). The distributions are shown along with that of a simulated
field with the same mean density of galaxies, but no galaxy clustering. In the
simulation, the redshift of galaxy formation is set at z$_f$=7,
and the cosmological parameters are  H$_0$=50~\kmsMpc and $\Omega_0$=1. Two
redshift bins show an excess of galaxies, at z$\sim$0.2 and at
z$\sim$0.9. These two concentrations in the redshift space  appear in both types of
distributions, whether z$_{phot}$ or  z$_{wm}$ is considered (see the two panels in Fig.~\ref{fig5}).  Moreover, the
shapes of the distributions do not change if a redshift
limit is imposed in {\it Hyperz}, for example at 1.0 or 1.5. Therefore, the galaxy excesses
at z$\sim$0.2 and z$\sim$0.9 are not related to the choice of the
redshift limit.

\begin{table*}[t]
\begin{center}
\begin{tabular}{|l|l|l|c|c|c|}
 \hline Model & Fitted parameters & Fixed parameters & $\chi^2$ & T$_d$&
$\cal{M}/\cal{L}_I$ \\
  \hline \hline  
L$_1$ &  r$_{cut}$(L)=51~kpc   $\sigma_L$=203 \kms &
$\epsilon_L$, $\theta_L$ & 113  & -- &7.5  \\
\hline  
L$_0$ &  r$_{cut}$(L)=51~kpc   $\sigma_L$=203 \kms
 $\theta_L$= 3\degr &
$\epsilon_L$  & $<$1  & $\sim 130$  &7.5  \\
\hline    
(L+M)$_1$   &  $\sigma_L$=198 \kms $\sigma_M$=135 \kms &
 $\epsilon_L$, $\theta_L$, r$_{cut}$(L)
 $\epsilon_M$, $\theta_M$, r$_{cut}$(M) & 5    &   --    &  7.1  \\
\hline    
(L+M)$_0$   &  $\sigma_L$=198 \kms $\sigma_M$=135 \kms $\theta_L$=-9\degr &
 $\epsilon_L$, r$_{cut}$(L)
 $\epsilon_M$, $\theta_M$, r$_{cut}$(M)  & $<1$ &   $\sim 111$  &  7.1  \\
\hline    
(L+M+C)$_1$     &   $\sigma_L$=189 \kms $\sigma_C$=718 \kms  &
 $\epsilon_L$, $\theta_L$, r$_{cut}$(L)
 $\epsilon_M$, $\theta_M$, r$_{cut}$(M), r$_{cut}$(C) &
  $<$1  & $\sim$ 100  & 6.5\\
\hline
\end{tabular}
\caption{\label{results}
Summary of the models studied. The quantity $\epsilon $ is the ellipticity of the object, $\theta $ is its PA,  $\sigma$  its velocity dispersion and  $\rm{r_{cut}}$ its cut radius. 
Column (1) indicates which components have been considered (galaxy L, 
galaxy M, cluster C) and the indice 0 or 1 represents the number of degree 
of freedom of the model.
Column (2) provides the derived values for the fitted parameters.
Column (3) summarizes the fixed parameters: in all cases the positions of the 
components were fixed to their observed values; the ellipticities
for galaxies L and M where fixed to their observed values; when fixed, 
$\theta_L$
is -23\degr and $\theta_M$ is -40\degr;
r$_{cut}$(L)= 51 kpc, r$_{cut}$(M)=20 kpc and r$_{cut}$(C)= 600 kpc.
Column (4) provides the $\chi^2$ value.
Column (5) gives the time-delay prediction expressed in days, for 
\ho=65~\kmsMpc $\Omega$=0.3 and $\Lambda$=0.7.
Finally, Column (6) displays the mass-to-light ratio value  $\cal{M}/\cal{L}_I$
computed within the Einstein radius for galaxy L
 (solar units, see section \ref{model}). 
}
\end{center}
\end{table*}

The color image on Fig. \ref{couleur} illustrates this result as well: the
galaxy population with violet colors corresponds to galaxies around  z=0.2 while the galaxy population with red color corresponds
to galaxies around  z=0.9.

As the galaxies contributing to the red population (redshift peak
around 0.9) are  also those which build up the overdensities seen in
Fig.~\ref{alfosc} and \ref{fig5bis}, we deduce that the cluster/group
located to the West of the quasar images is indeed at redshift
z=0.9$^{+0.10}_{-0.25}$ (3$\sigma$ error bars). We cannot exclude that there might be a real difference between the redshift of the main lensing galaxy (z=0.71) and the redshift of the cluster/group. However, we consider in the modelling (see section \ref{model}) that they are both at z=0.71, as this is still consistent with the error--bars.  


\section{Lens modelling}\label{model}

Analytical modelling of \SBS\, was performed using the {\it Lenstool}
software developed by Kneib et al. (1993). This software computes 
the lens mass distribution necessary to reproduce the
configuration and flux ratios of multiply imaged quasars. In the present case,
multiple component lenses can be used, each one being a truncated
Pseudo-Isothermal-Elliptical-Mass-Distribution (PIEMD, Kassiola \&
Kovner 1993, Kneib et al. 1996, Hjorth \& Kneib 2002). Given the
observational constraints available we explored a number of lens
configurations, involving a single lensing galaxy (model L), two
galaxies (model L+M), and two galaxies plus a galaxy cluster (model
L+M+C). Obviously, the galaxy cluster is represented by a circular PIEMD
mass distribution.

The constraints available are the positions of the quasar images and
their flux ratio in the F160W band (Table \ref{tab1}), the positions of galaxies L and M and their PA ($\theta$) and ellipticities ($\epsilon$)
with error--bars given in section
\ref{results} and the position of the cluster. The error-bars on PA take 
into account that the misalignment between dark-matter halo and light of a 
galaxy could be of 10$\degr$ (Kochanek 2001), so that the galaxy mass 
distribution could have an orientation different from that traced by 
the light.\\

 Given the results and uncertainties from
the photometric redshift analysis (see section 3 and 4), a same redshift can be considered for all model components. We take this redshift to be that of
the main lensing galaxy, z=$0.71$.  
\subsection{Model L}
We first run a model considering only the lensing galaxy L. The fitted
parameters, $r_{cut}(L)$ and $\sigma_L$
 are given in Table~\ref{results}. 
The quantities $\epsilon_L$ and $\theta_L$, as well as the position of the lens,
 are fixed to their observed values within their respective error bars.
In this case, neither the flux ratio nor the relative position of the quasar 
images are recovered, as shown by the large value of the $\chi^2$.
This strengthens the suspicion that galaxy L is not the only intervening lens.

\subsection{Model L+M}
Therefore, in a second step we consider that
galaxy M (the nearest galaxy to galaxy L) contributes to the total lensing 
potential as well. Again the quantities $\epsilon_L$, $\epsilon_M$, 
$\theta_L$, $\theta_M$, and the positions  of the two galaxies are fixed to 
their observed values within their respective error bars.
In this case we recover the flux ratio, but not the geometrical configuration 
of the quasar images.

\subsection{Model L+M+C}
Therefore, additional mass is necessary to fit both the configuration and 
flux ratio of the quasar images. A contribution from a galaxy cluster is
 proposed and introduced in the model. We first center the
cluster between the two peaks detected in the galaxy density map (see
Fig.~\ref{alfosc}), but we fail to obtain an acceptable $\chi^2$
fit to the data. The best $\chi^2$ values are obtained for a centering
 of the cluster component on the NW peak of the galaxy overdensity.
Doing this, we obtain the results summarized in the last line of Table
\ref{results}. Adding the SW peak of the galaxy overdensity as a forth PIEMD still provides a good fit. Indeed, being further away, its contribution is minor in the lens model and its unique impact is to reduce the time--delay by a few days.\\

\subsection{Inferences from the lens modelling}
The mass to light ratio $\cal{M}/\cal{L}_I$ of galaxy L computed
within the Einstein radius (r$_E \sim$156 kpc or 0.7\arcsec) for the
 different models is given in Table~\ref{results}. 
The luminosity $\cal{L}_I$ was determined directly from the observed
F160W photometry, as the F160W filter for a galaxy at $z\sim 0.9$
roughly correspond to the I-band filter at z=0, and thus no color
correction was applied. For an elliptical galaxy $\cal{L}_B/\cal{L}_I \sim $1.9
therefore the derived $\cal{M}/\cal{L}_B$ for the galaxy L, 
is consistent with the range of usual $\cal{M}/\cal{L}_B$
ratios found in lens systems (Keeton et al 1998).\\

The time--delay  predictions given in Table
\ref{results} were computed for the various models of the lens, 
using \ho=65~\kmsMpc $\Omega$=0.3 and $\Lambda$=0.7. 
Changing the cosmology to $\Omega$=1
and $\Lambda$=0, with the same value for \ho, decreases the time--delay to
85 days for the L+M+C model. Changing the value of \ho\, from 
65~\kmsMpc to 50~\kmsMpc with $\Omega$=0.3 and $\Lambda$=0.7, increases the time--delay from 100 days to 125 days for 
the L+M+C model.

\section{Concluding remarks}\label{sec4}

We have studied the doubly imaged quasar \SBS\, using both HST and
ground based data, in order to map the projected mass distribution of
the lensing system at small and large scales.

First, by deconvolving the HST/NICMOS data, we have accurately
measured the shape and brightness of the main lensing galaxy. We detect a
weak extension to the NW of the lensing galaxy (Fig.~1).  This feature
does not seem to be a residual of the deconvolution process. If it is a genuine
feature, it may have some influence on the lens model, considering its
proximity to image A of the quasar. However until it has been confirmed, we ignore it for the modelling.


Second, we have mapped the relative surface density of faint galaxies
around the quasar images and identified an overdensity of objects to the
West. Photometric redshifts were estimated
for the main lensing galaxy L as well as for all the galaxies in the
7.5~(\arcmin)$^2$ surrounding the lensed quasar. 
 This analysis reveals a concentration in redshift space around a mean value z=0.9$^{+0.10}_{-0.25}$ and confirms
the 3D--reality of the observed projected overdensity of objects: we consider it to be a galaxy cluster. 
As a consequence, it is probable that the main lensing galaxy L (measured spectroscopic redshift z=0.71) is a member of this cluster (measured photometric redshift z=0.9$^{+0.10}_{-0.25}$). Following this argument and considering that the spectroscopic redshift measurement is more accurate, we have built our lens model with a unique redshift value,  z=0.71, for the main lensing galaxy, for galaxy M and for the cluster.


Finally, the presence of a galaxy cluster to the W of \SBS\, is
also suggested by the lensing model analysis which does require such 
a contribution. If not taken into account, it is impossible to reproduce 
the observed configuration and flux ratio of the double quasar images.
This argument is valid as long as light traces mass, which seems to be the 
case for most lensing galaxies (Kochanek 2001).

An
independent and complementary way of probing the line of sight mass
distribution is to measure the time--delay induced by the lens system (also necessary to constrain \ho). 
Therefore, it is interesting to predict the time-delay corresponding to
the lens model derived from the lensed quasar image configuration. We have made time-delay predictions for the 3 different lens models analysed in this paper, to be compared with the observed time-delay, when it will be available.

Yet, further detailed observations are needed to improve the understanding of the lensing system and enable to use
it as a cosmological probe.   \SBS\, will be observed by the X--ray satellite {\it
Chandra} (PI: G.Garmire): this observation should confirm the
existence of the cluster in a way similar to   
the quadruply lensed quasar RX~J0911.4+0551 (Morgan et al. 2001).
Deep Keck/Gemini multi-object spectroscopy would be invaluable to
settle the redshift of the detected cluster/group and provide an estimate of its mass.

\begin{acknowledgements}

It is a pleasure to thank Jens Hjorth and Andreas Jaunsen for 
providing the lensing galaxy redshift prior to publication.  
We also acknowledge constructive remarks from the referee, David Rusin.
C\'ecile Faure acknowledges support from an ESO studentship in Santiago. F.C. is supported by Chilean grant  FONDECYT 3990024, by the European
  Southern
  Observatory,   and   by  a Marie   Curie   grant  MCFI-2001-00242.   Two
  collaborative  grants between  Chile  and France  are also  gratefully
  acknowledged:  ECOS/CONICYT   CU00U05  and  CNRS/CONICYT   8730.  
Jean--Paul Kneib thanks
CNRS for support as well as ESO for a productive visit in Santiago. 

\end{acknowledgements}

\end{document}